# Redetermination of the second-order nonlinear optical coefficients of $BaMgF_4$

**Shoichi Ui,[1] Simon J. Herr,[2] Gaetano G. M. Bonetti,[1] Ingo Breunig,[2,3] Frank Kühnemann,[2,4] Matthias Bickermann,[1] and Hiroki Tanaka[1,*]**

[1]*Leibniz-Institut für Kristallzüchtung (IKZ), Max-Born-Str. 2, 12489 Berlin, Germany*
[2]*Fraunhofer-Institut für Physikalische Messtechnik IPM, Georges-Köhler-Allee 301, 79110 Freiburg, Germany*
[3]*Laboratory for Optical Systems, Department of Microsystems Engineering – IMTEK, University of Freiburg, Georges-Köhler-Allee 102, 79110 Freiburg, Germany*
[4]*University of Freiburg, Institute of Physics, Freiburg, Germany*
**hiroki.tanaka@ikz-berlin.de*

**Abstract:** $BaMgF_4$ (BMF) has the potential to serve as a platform material for vacuum-ultraviolet (VUV) nonlinear optics owing to its ferroelectricity and high UV transparency down to 130 nm. However, the second-order nonlinear optical coefficients reported for BMF span approximately one order of magnitude, hindering a reliable assessment of its potential for nonlinear optical applications. Here, we redetermined these coefficients by combining Maker fringe, wedge, and single-pass second-harmonic-generation measurements at a fundamental wavelength of 1064 nm. Our results establish that all the second-order nonlinear optical coefficients of BMF lie in the range of 0.01–0.04 pm/V. We determined $d_{31}$ to be 0.035 pm/V, which is among the largest coefficients of BMF, while it has been underestimated relative to the other coefficients in previous studies. Although $d_{31}$ is not the largest coefficient, the use of its long corresponding coherence length is beneficial for quasi-phase matching, particularly for VUV generation, where the fabrication of lower-order quasi-phase-matching structures is crucial.

## 1. Introduction

Coherent light sources in the vacuum-ultraviolet (VUV) spectral range below ~200 nm have recently attracted growing interest owing to industrial and scientific applications, such as semiconductor wafer inspection [1], Raman spectroscopy [2,3], and angle-resolved photoemission spectroscopy (ARPES) [4]. In addition, a 186.75-nm light source is required for a recently proposed $Sn^{2+}$-based optical atomic clock [5], and a 148.4-nm light source for a $^{229}Th$ optical nuclear clock [6]. These next-generation precision clocks have motivated the development of narrow-linewidth laser sources at these wavelengths.

Several nonlinear materials have been investigated for these applications, such as $SrB_4O_7$ (SBO) and $KBe_2BO_3F_2$ (KBBF). SBO has a nonlinear optical (NLO) coefficient as large as 1.5 pm/V [7]. However, since SBO is not ferroelectric, a periodically-poled structure for quasi-phase-matching (QPM) cannot be introduced by applying an external electric field. Wavelength conversion in SBO has instead been achieved by random QPM based on the twinned domain structure formed in grown crystals [8,9], and the wavelength conversion to 148.4 nm has been demonstrated [10]. However, controlling the formation of such domain structures is challenging. KBBF satisfies birefringent phase-matching (BPM) for second-harmonic generation (SHG) down to 161 nm [11,12] and for sum-frequency generation down to 149.8 nm [13], close to its absorption edge. At 193 nm, more than 15 mW output has been achieved in continuous-wave operation [14]. The drawbacks of KBBF, however, are its toxic beryllium content and the need for a prism-coupling device to access the phase-matching angle, because of the challenge of growing sufficiently thick crystals that can be cut at the required angle. Many fluorooxoborate crystals have been studied to overcome these limitations [15]. A representative example, $NH_4B_4O_6F$ (ABF), can be grown up to centimeter size and enables BPM down to 158.9 nm [16].

SHG based on barium magnesium fluoride ($BaMgF_4$, BMF) is a promising route for VUV wavelength conversion. BMF is a non-centrosymmetric, ferroelectric crystal, known for its wide transparency range extending from 130 nm to 13 µm [17,18]. Although the second-order NLO coefficients of BMF are relatively small, its ferroelectricity enables periodic domain inversion, while its high damage threshold and UV resistance allow operation at high optical intensities, enhancing the conversion efficiency. The damage threshold was reported to be at least ~$10^9$ W/cm$^2$ at 1064 nm [19], and its high resistance to UV irradiation was demonstrated by the absence of transmission degradation, after irradiation with more than $1.1\times10^9$ pulses at 157 nm (10 ns, 2 mJ/cm$^2$ per pulse) [20]. These properties make BMF a promising platform for all-solid-state coherent light sources via QPM, particularly in the VUV spectral range.

Wavelength conversion to 148.4 nm for $^{229}$Th optical nuclear clocks has recently been achieved with a periodically-poled BMF [21], but with a limited conversion efficiency due to the use of 9$^{th}$-order QPM; thus, further optimization is required to improve the conversion efficiency.

Accurate knowledge of the NLO coefficients is essential for optimizing the conversion efficiency and designing efficient QPM devices, as the SHG output scales quadratically with the effective NLO coefficients. BMF is an orthorhombic crystal that belongs to the point group mm2. The twofold-rotation polar axis is the crystallographic *c*-axis ($c$=5.821 Å), and the *a*- and *b*-axes are assigned according to the convention of $a < b$ ($a$ =4.126 Å, $b$ = 14.518 Å) [22]. This axis assignment has consistently been adopted in previous studies of nonlinear optical properties of BMF crystals [19,23,24]. The second-order NLO coefficient tensor has five non-zero elements. Using the contracted matrix $d_{il}$ ($d_{il}=1/2\chi_{jl}^{(2)}$), the second-order nonlinear polarization for SHG is expressed as

$$\begin{bmatrix} P_x(2\omega) \\ P_y(2\omega) \\ P_z(2\omega) \end{bmatrix} = \varepsilon_0 \begin{bmatrix} 0 & 0 & 0 & 0 & d_{15} & 0 \\ 0 & 0 & 0 & d_{24} & 0 & 0 \\ d_{31} & d_{32} & d_{33} & 0 & 0 & 0 \end{bmatrix} \begin{bmatrix} E_x^2(\omega) \\ E_y^2(\omega) \\ E_z^2(\omega) \\ 2E_y(\omega)E_z(\omega) \\ 2E_z(\omega)E_x(\omega) \\ 2E_x(\omega)E_y(\omega) \end{bmatrix}. \quad (1)$$

Here, $x,y,z \equiv a,b,c$. The subscripts $\omega$ and $2\omega$ denote the fundamental and the second-harmonic (SH) frequencies, respectively. Assuming Kleinman symmetry, $d_{31} = d_{15}$ and $d_{32} = d_{24}$. Table 1 summarizes three previous studies that reported the NLO coefficients of BMF, $d_{il}$, at a fundamental wavelength of 1064 nm together with the corresponding coherence lengths, $l_c$ [19,23,24]. The reported NLO coefficients, however, show a substantial discrepancy between the two studies from the 1970s [19,23] and the later study in 2012 [24].

In this work, using high-quality BMF single crystals grown by the Czochralski technique, we determined its NLO coefficients and coherence lengths by combining wedge and Maker fringe measurements, using $d_{11}$ (0.30 pm/V) of α-$SiO_2$ (α-quartz) as the reference for absolute scaling of the NLO coefficients, rather than $d_{31}$ (4.35 pm/V) of $LiNbO_3$ used in [24]. The resulting coefficient ratio of $d_{31}/d_{32}$ was further supported by independent single-pass SHG experiments. Our results confirm that the NLO coefficients of BMF are in the range of 0.01–0.04 pm/V, which is consistent with the reports from the 1970s [19,23], and show that the coefficients $d_{31}$ and $d_{15}$ were underestimated.

**Table 1. Previously reported second-order NLO coefficients of BMF. In [19] and [23], the NLO coefficients were reported in the relative scale with their references, and the uncertainties were reported in percentage. Here, the NLO coefficients are calculated using the representative values of $d_{11}$ = 0.30 pm/V for α-quartz and $d_{36}$ = 0.39 pm/V for $KH_2PO_4$ (KDP) [25]. The uncertainties are also calculated into absolute value. In [24], the NLO coefficients were scaled using $LiNbO_3$ ($d_{31}$=4.35 pm/V) as reference.**

| Ref. | [19] | [23] | [24] |
|---|---|---|---|
| NLO reference [pm/V] | α-quartz $d_{11}$=0.30 | KDP $d_{36}$=0.39 | $LiNbO_3$ $d_{31}$=4.35 |
| $d_{31}$ [pm/V] | 0.021±0.004 | 0.022±0.005 | 0.15 |
| $d_{32}$ [pm/V] | 0.039±0.008 | 0.033±0.004 | 0.36 |
| $d_{33}$ [pm/V] | 0.015±0.003 | 0.009±0.001 | 0.12 |
| $d_{15}$ [pm/V] | 0.021±0.004 | - | - |
| $d_{24}$ [pm/V] | 0.021±0.004 | 0.024±0.004 | - |
| $l_{c31}$ [μm] | ∞ | 665±133 | 380 |
| $l_{c32}$ [μm] | 9.6 | 10.9±0.2 | 11 |
| $l_{c33}$ [μm] | 29.3 | 35.7±1.8 | 35 |
| $l_{c15}$ [μm] | 23.3 | 25.83* | - |
| $l_{c24}$ [μm] | 170 | 222±33 | - |

*Calculated value from refractive indices.

## 2. Rotational Maker fringe and wedge measurements

### 2.1 Crystal growth and sample preparation

We grew several BMF single crystals by the Czochralski technique and prepared *a*- and *b*-cut plane-parallel as well as wedged samples, used for measurements with rotational Maker fringe and wedge techniques. The detailed growth procedure is described in [26]. Note that our crystals grown under the same conditions were also used for the 148-nm generation [21]. As a reference for the absolute NLO coefficients, we used a Y-cut (cut perpendicularly to $[01\bar{1}0]$-axis) α-quartz sample with the NLO coefficient $d_{11}$ known to be 0.30 pm/V [25].

### 2.2 Rotational Maker fringe measurement

Figure 1 illustrates the experimental setup. As a pump source, we used a Q-switched laser at 1064 nm, delivering 1-ns pulses with a pulse energy of ≈75 μJ at a repetition rate of 1 kHz (CryLaS, DSS1064-Q3). Its output was focused onto a sample mounted on motorized translation and rotation stages. The beam diameter was measured to be 360 μm × 420 μm at the sample position. The polarization of the fundamental light was controlled with a half-wave plate and a polarizer. The SHG signal of a selected polarization was detected using a photomultiplier tube (PMT, Hamamatsu, R7207-01) after the fundamental light was suppressed by a harmonic separator and a bandpass filter (BPF). To measure and compare SH signals at substantially different intensities, neutral density (ND) filters with known transmittances were used to attenuate the SH signals when necessary. To compensate for the peak-power fluctuation of the fundamental pulses, the SH signal from a β-$BaB_2O_4$ (BBO) crystal was simultaneously monitored using a photodetector (PD). All optics, except for BMF samples, were anti-reflection coated at the fundamental or SH wavelengths. The signals from PD and PMT were respectively averaged in boxcar integrators (Stanford Research Systems, SR250) for 100 pulses to improve the signal-to-noise ratio. The polarization angle of the fundamental light was chosen to isolate SH signals from different NLO coefficients. Figure 2 shows the BMF sample orientation and polarization setting in the Maker fringe measurements for the five NLO coefficients.

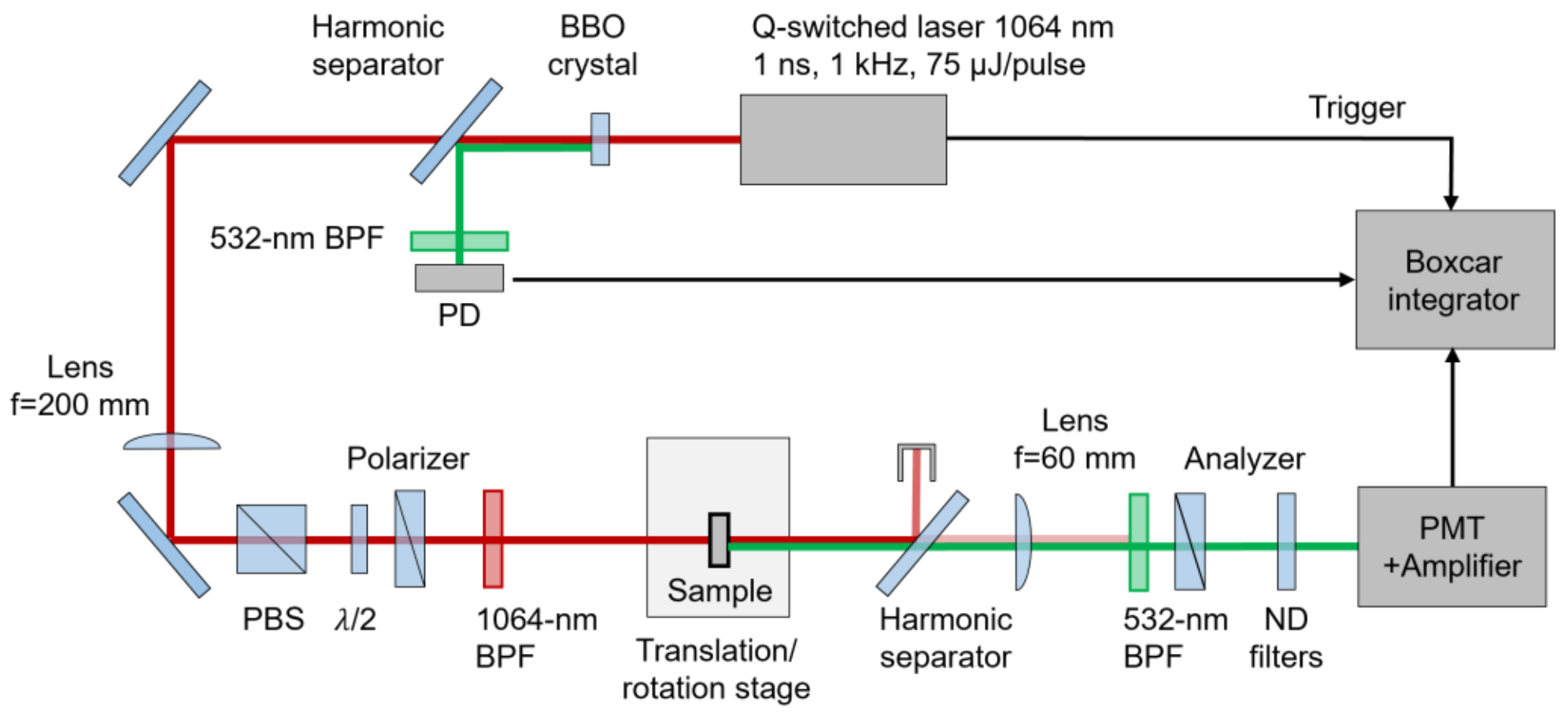

Fig. 1 Experimental setup. BPF: bandpass filter, PD: photo detector, PBS: polarizing beam splitter, λ/2: half-wave plate, ND filter: neutral density filter, PMT: photomultiplier tube

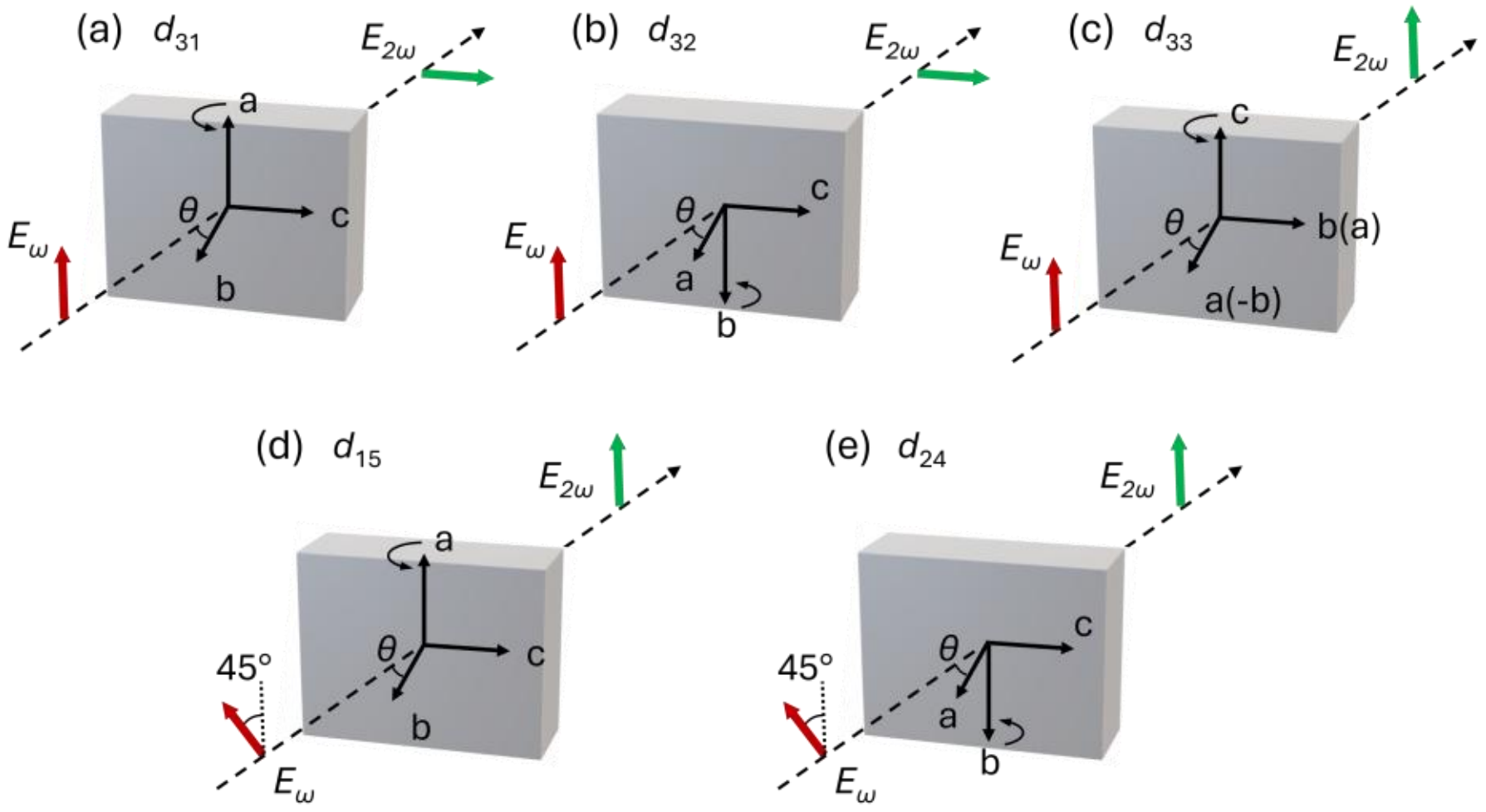

Fig. 2 Experimental configuration in the Maker fringe measurements for (a) $d_{31}$, (b) $d_{32}$, (c) $d_{33}$, (d) $d_{15}$, (e)$d_{24}$ coefficients of BMF. For $d_{33}$, both *a*- and *b*-cut samples can be used.

The SHG intensity, $I_{2\omega}$, is given by [23]

$$I_{2\omega}(\theta) = d_{il}^2 f(\theta) I_\omega^2 \sin^2\Psi, \tag{2}$$

where $\theta$ is the angle of incidence, $f(\theta)$ is the envelope function (see Appendix 1), and Ψ is half of the phase difference between the fundamental and SH waves, given by

$$\Psi = \frac{2\pi L}{\lambda_\omega}(n_\omega \cos\theta'_\omega - n_{2\omega}\cos\theta'_{2\omega}). \tag{3}$$

Here, $L$ is the sample thickness, $\lambda_\omega$ is the fundamental wavelength, $n$ is the refractive index, and $\theta'$ is the refraction angle. The detailed expressions of SHG intensity for isotropic and uniaxial crystals are given in [27], which we used to analyze the result of α-quartz reference sample, and those for biaxial crystals in point group mm2 are given in [23]. Equations (2) and (3) allow us to calculate the relative NLO coefficients by comparing the envelope intensities at the normal incidence, $I_{\text{env}}(0) = d_{ij}^2 f(0) I_\omega^2$, as

$$\frac{d_{il,\mathrm{BMF}}}{d_{11,\alpha\text{-quartz}}} = \sqrt{\frac{I_{\mathrm{env,BMF}}(0)}{f_{\mathrm{BMF}}(0)}} \Bigg/ \sqrt{\frac{I_{\mathrm{env},\alpha\text{-quartz}}(0)}{f_{\alpha\text{-quartz}}(0)}}. \quad (4)$$

To validate our experimental setup, we compared the two representative NLO references, $d_{11}$ of α-quartz and $d_{36}$ of $KH_2PO_4$ (KDP) crystals. We obtained an envelope intensity ratio of $I_{\mathrm{env},\alpha\text{-quartz}}/I_{\mathrm{env,KDP}} = 1.7 \pm 0.1$, resulting in a NLO coefficient ratio of $d_{11,\alpha\text{-quartz}}/d_{36,\mathrm{KDP}} = 0.77 \pm 0.03$. These values are consistent with previous reports [25,28].

The obtained Maker fringes are shown in Fig. 3. As reported in [23], we observed the type-I phase-matching in the *bc*-plane for the $d_{31}$ measurement (Fig. 3(a)), at an incident angle $\theta$ of 13.8°. As the SHG intensity at the phase-matching peaks in $d_{31}$ measurement was too high to be measured with PMT, we restricted the rotation angle to $|\theta| < 10°$ (Fig. 3(a)). To observe a sufficient number of fringes in the limited angle range, we used a ~10-mm thick sample.

In the $d_{15}$ measurement (Fig. 3(d)), the rotation angle was also restricted to $|\theta| < 10°$ to avoid the detection of type-I phase-matching SH signal due to $d_{31}$. The intense phase-matched signal, despite being orthogonally polarized to the signal originating from $d_{15}$ (cf. Figs. 2(a) and (d)), could not be completely eliminated even with three polarizers. Despite the restricted rotation angle, we still observed an unexpected Maker fringe shape as well as an offset. We attributed this to the interference with the SH signal from other NLO coefficients. In our analysis, we subtracted the offset and fitted the theoretical equation to the experimental data in the range of $|\theta| < 5°$. Note that a similar difficulty in measuring $d_{15}$ due to the phase-matching peak was also reported in [23].

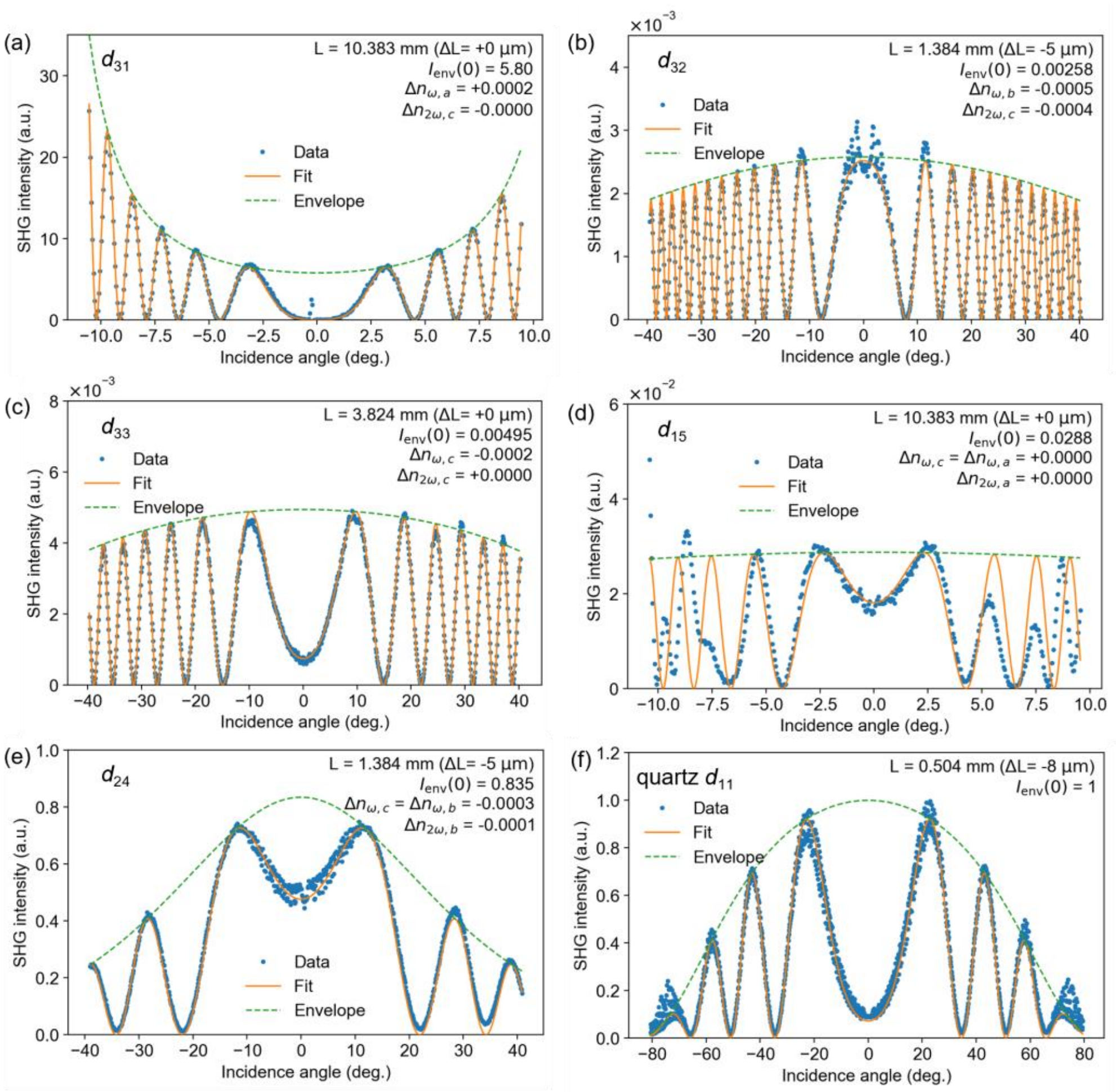

Fig. 3 Maker fringes of BMF for the NLO coefficients of (a)$d_{31}$ (b)$d_{32}$ (c)$d_{33}$ (d)$d_{15}$ (e)$d_{24}$. The vertical scale is normalized to the envelope peak of SHG intensity obtained from the reference α-quartz sample shown in (f). For each measurement, the best fit results are denoted in the respective figure: the sample thickness $L$, with the correction $\Delta L$ to the measured value, the envelope intensity at normal incidence $I_{\mathrm{env}}(0)$, and corrections for refractive indices $\Delta n$.

For each Maker fringe, we first calculated the theoretical fringes using the refractive indices based on the reported Sellmeier equations [18]. However, the calculated fringes did not fit the experimental ones, suggesting the limited accuracy of the reported Sellmeier equations. The discrepancy was also observed in the type-I phase-matching angle in the *bc*-plane. Phase-matching was observed at $\theta$=13.8°, whereas the Sellmeier equations predict it at $\theta$=17.2°, corresponding to a refraction angle of $\theta'$=11.7° inside the crystal. Therefore, we introduced refractive-index correction terms, $\Delta n_{\omega}$ and $\Delta n_{2\omega}$, as additional fitting parameters to be added to the refractive indices from the Sellmeier equations. The applied corrections are indicated in each panel of Fig. 3. These correction terms were added to each principal refractive index and were on the order of $10^{-4}$, suggesting that the precision of the reported Sellmeier equations is limited to this order. Furthermore, we slightly varied the sample thickness by a few micrometers for better fitting. The Maker fringes, particularly for $d_{32}$ and $d_{24}$ of BMF as well as $d_{11}$ of α-quartz, are affected by multiple reflections of the fundamental light between the two parallel sample surfaces [29,30]. The resulting interference oscillations were considered in estimating the uncertainties. The systematic overestimation of the NLO coefficients due to multiple reflections was negligible, as it is reported to be less than 5% for crystals with refractive indices smaller than 2, including BMF [31].

### *2.3 Determination of the coherence lengths by wedge technique*

The determination of NLO coefficients from the rotational Maker fringe results requires the accurate knowledge of coherence lengths for normal incidence, $l_c(\theta=0)$, which is defined as a propagation length over which the phase difference between the fundamental and SH waves, $2\Psi$, becomes $\pi$, given as

$$l_c(\theta = 0) = \frac{\lambda_\omega}{4|n_\omega - n_{2\omega}|}. \tag{5}$$

The envelope function of the Maker fringes $f(\theta)$ in Eqs. (2) and (4) involves the term $1/(n_\omega^2 - n_{2\omega}^2)^2$ (see Appendix 1), which is inversely proportional to $l_c^2$. Accordingly, the calculated NLO coefficients are affected by the accuracy of the coherence lengths, as $d_{il} \propto l_c^{-1}$. However, the coherence lengths calculated from Eq. (5) could be highly sensitive to the refractive index errors, particularly when $|n_\omega - n_{2\omega}|$ is on the order of $10^{-4}$, comparable to the precision of the reported Sellmeier equations [18].

Therefore, we experimentally determined the coherence lengths by the wedge technique. The results are shown in Fig. 4. The obtained fringes were fitted using Eq. (2) for normal incidence ($\theta = \theta' = 0$), with $n_\omega - n_{2\omega}$ as a fitting parameter. For $d_{31}$ and $d_{24}$, the corresponding coherence lengths were longer than 200 µm. We therefore used three and two samples, respectively, with different thicknesses but the same wedge angle of 0.54±0.01°, to observe sufficient data enabling reliable sinusoidal fitting. Only for $d_{15}$, we could not obtain a periodic $\sin^2\Psi$-shaped curve. This is attributable to the SHG signal from other NLO coefficients. Since the coherence length for $d_{15}$ could not be determined from the fringe, we adopted a value calculated from the refractive indices from the reported Sellmeier equations [18]. As the corresponding $|n_\omega - n_{2\omega}|$ is on the order of $10^{-2}$, the coherence length for $d_{15}$ can be calculated with sufficient accuracy from the reported Sellmeier equations. Some fringes exhibit unexpected spikes, especially for $d_{24}$, which might be attributed to surface defects, but further investigation is required to clarify their origin. Note that no grain boundaries were observed in the samples used in the experiments.

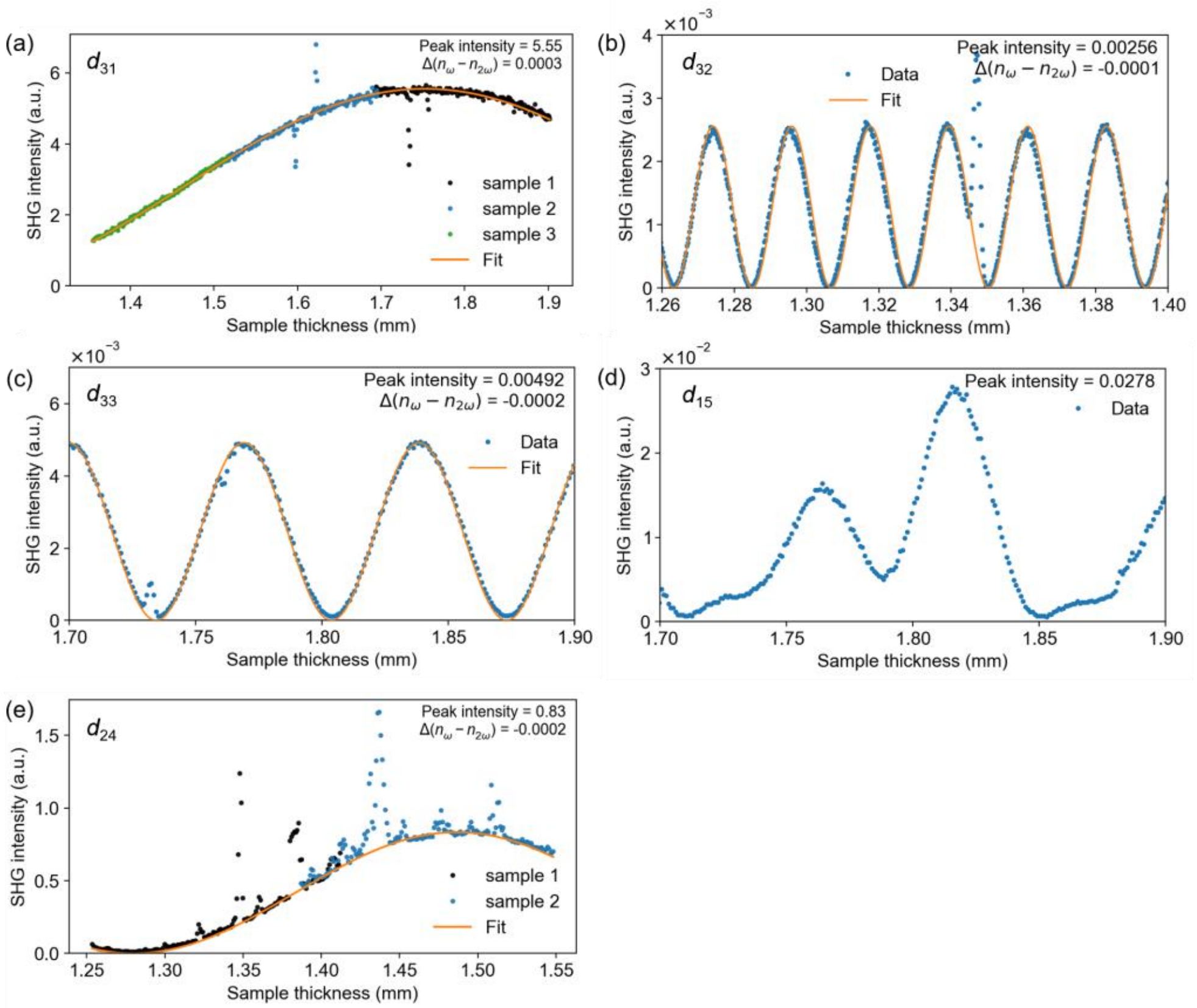

Fig. 4 The results of wedge measurements for (a)$d_{31}$ (b)$d_{32}$ (c)$d_{33}$ (d)$d_{15}$ (e)$d_{24}$. The vertical scale is normalized to the peak envelopeintensity of the rotational Maker fringe measurement measured for the reference α-quartz. The values of the peak intensity and $\Delta(n_\omega - n_{2\omega})$ obtained for the best fit are indicated in each figure.

The dependence of the coherence lengths on the errors of $n_\omega - n_{2\omega}$, calculated from Eq. (5), is shown in Fig. 5 as the black solid curves, together with the previously reported coherence lengths [19,23] and our results. The coherence lengths for $d_{31}$ and $d_{24}$ are shown to be particularly sensitive to the errors in refractive indices.

### *2.4 Determination of the second-order nonlinear optical coefficients*

Using the determined coherence lengths, we calculated the NLO coefficients according to Eq. (4). The results are summarized in Table 2. The uncertainties in the NLO coefficients were calculated from those in the coherence lengths and the envelope intensities. The coefficient $d_{15}$ is identified to be the largest. This value is consistent with the equality $d_{31}$=$d_{15}$ under Kleinman symmetry. The five coefficients are fairly consistent with those reported in the 1970s [19,23], while they are significantly different from the latest report [24]. We discuss this difference in Section 4. It is noteworthy that the value of $d_{31}$ is found to be ~1.5 times larger than reported in the earlier two papers and is comparable to $d_{32}$, which was previously reported to be the largest NLO coefficient of BMF. This result is further examined in the following section.

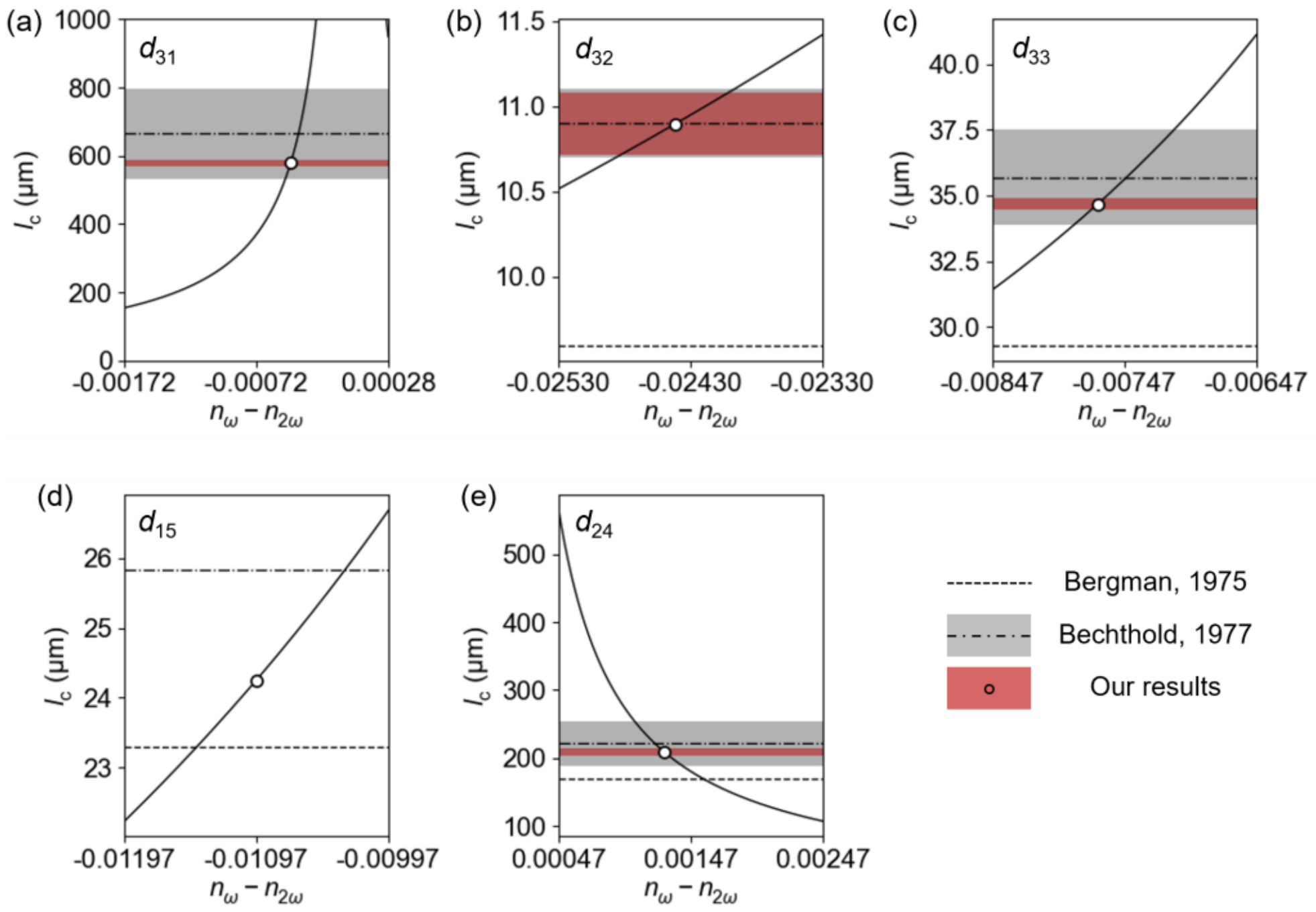

Fig. 5 Dependence of BMF coherence length $l_c(\theta=0)$ on the value of $n_\omega - n_{2\omega}$ (solid line) for (a)$d_{31}$ (b)$d_{32}$ (c)$d_{33}$ (d)$d_{15}$ (e)$d_{24}$. The horizontal axis spans ±0.001 around the value calculated from the reported Sellmeier equations [18]. The previously reported coherence lengths [19,23] and those obtained in this work are also shown, with the colored bands indicating uncertainties.

**Table 2. The Second-order NLO coefficients of BMF determined in this work. The coherence length and its uncertainty for $d_{15}$ are calculated using the reported Sellmeier equations** [18]**, assuming a deviation of ±0.001 in the value of $n_\omega - n_{2\omega}$ as shown in Fig. 5(d).**

| $d_{il}$ | NLO coefficient [pm/V] | Coherence length [μm] |
|---|---|---|
| $d_{31}$ | 0.035±0.002 | 580±10 |
| $d_{32}$ | 0.038±0.002 | 10.9±0.2 |
| $d_{33}$ | 0.0115±0.0004 | 34.6±0.2 |
| $d_{15}$ | 0.040±0.004 | 24±2 |
| $d_{24}$ | 0.024±0.001 | 210±5 |

## 3. Comparison of $d_{31}$ and $d_{32}$ via single-pass SHG

Contrary to the three previous reports on the NLO coefficients of BMF, which reported $d_{31}$ to be approximately half of $d_{32}$, our results suggest that $d_{31}$ is comparable to $d_{32}$. To further validate our results, we compared these two NLO coefficients via single-pass SHG experiments, under the type-I BPM in the *bc*-plane ($d_{eff} = d_{31}\cos\theta'_{PM}$, where $\theta'_{PM}$ is the phase-matching angle in the crystal) and under QPM using $d_{32}$ without any temperature control. The measurements were performed in a similar setup shown in Fig. 1, but the generated SH signal was detected using a power meter instead of the PMT. We also increased the repetition rate of the Q-switched laser to 2.5 kHz for higher average power, and reduced the beam spot diameter to 70 μm × 110 μm for higher conversion efficiencies. The sample for BPM was *b*-cut and 9.66 mm in thickness, while the one for QPM was a 5-mm-long *a*-cut sample with a poled length of 4.01 mm. The periodic domain inversion was performed using a calligraphic poling technique [26]. The sample had 183 cycles of inverted domains with a nominal period of 21.896 μm and ~40% duty ratio. The period, corresponding to $2l_{c32}$, was calculated from the reported Sellmeier equations [18]. The microscope image of the fabricated periodically-poled BMF is shown in Fig.6(a).

Figure 6(b) shows the average SHG power versus the incident fundamental laser power for BPM and QPM. As expected, the SHG output power increased quadratically with the incident fundamental power. Although we aimed QPM at normal incidence, the QPM peak was observed at an incident angle of ~9°, indicating the domain period being not exactly equal to the actual $2l_{c32}$. This result suggests the required poling period of 21.78 μm, corresponding to a shift in refractive indices of only ~$10^{-4}$, which falls within the error bar in Fig. 5(b). Thus, we conclude that the shift was due to the limited accuracy of the Sellmeier equations used. The peak from BPM was observed at an incident angle of 13.8°.

The SHG output power follows the Boyd-Kleinman equation [32], given by

$$P_{2\omega} = \frac{2\omega^2 d_{\mathrm{eff}}^2 L k_\omega}{\pi \varepsilon_0 c^3 n_\omega^2 n_{2\omega}} P_\omega^2 h(B,\xi), \tag{6}$$

where $P_{\omega,2\omega}$ are the peak power of the fundamental and SH wave, $k_\omega$ is the wavenumber of the fundamental, and $h(B,\xi)$ is the Boyd-Kleinman factor. Under loose focusing conditions, this factor is approximated as

$$h(B,\xi) \simeq \xi(1 - t^2/12 + t^4/120 - t^6/1344 + \cdots). \tag{7}$$

Here $B = \rho\sqrt{Lk_\omega}/2$, $\xi = L/b$, $b = r^2 k_\omega$, and $t = 2B\sqrt{2\xi}$, where $\rho$ is the walk-off angle, $L$ is the nonlinear interaction length, corresponding to the crystal thickness for BPM or the poled length for QPM, and $r$ is the beam radius in the plane of spatial walk-off. The effective NLO coefficient for the BPM is $d_{\mathrm{eff}} = d_{31}\cos\theta'_{\mathrm{PM}}$, whereas that for QPM is scaled by a factor of $S$, such that $d_{\mathrm{eff}} = d_{32}S$. Assuming perfect periodicity with a duty ratio of $D$, $S$ is $2\sin(\pi D)/\pi$ [33]. The walk-off angles between the fundamental and the SH wave at the corresponding angles of incidence were calculated to be 0.22° for BPM and 0.16° for QPM, respectively.

Taking these effects into account, the ratio of $d_{31}/d_{32}$ from the single-pass SHG measurement was calculated to be ≈0.84, in fairly good agreement with the ratio obtained from the Maker fringe and the wedge measurements: 0.92±0.07. These values are clearly larger than the ratios from previously reported coefficients: 0.53 [19], 0.67 [23], and 0.41 [24].

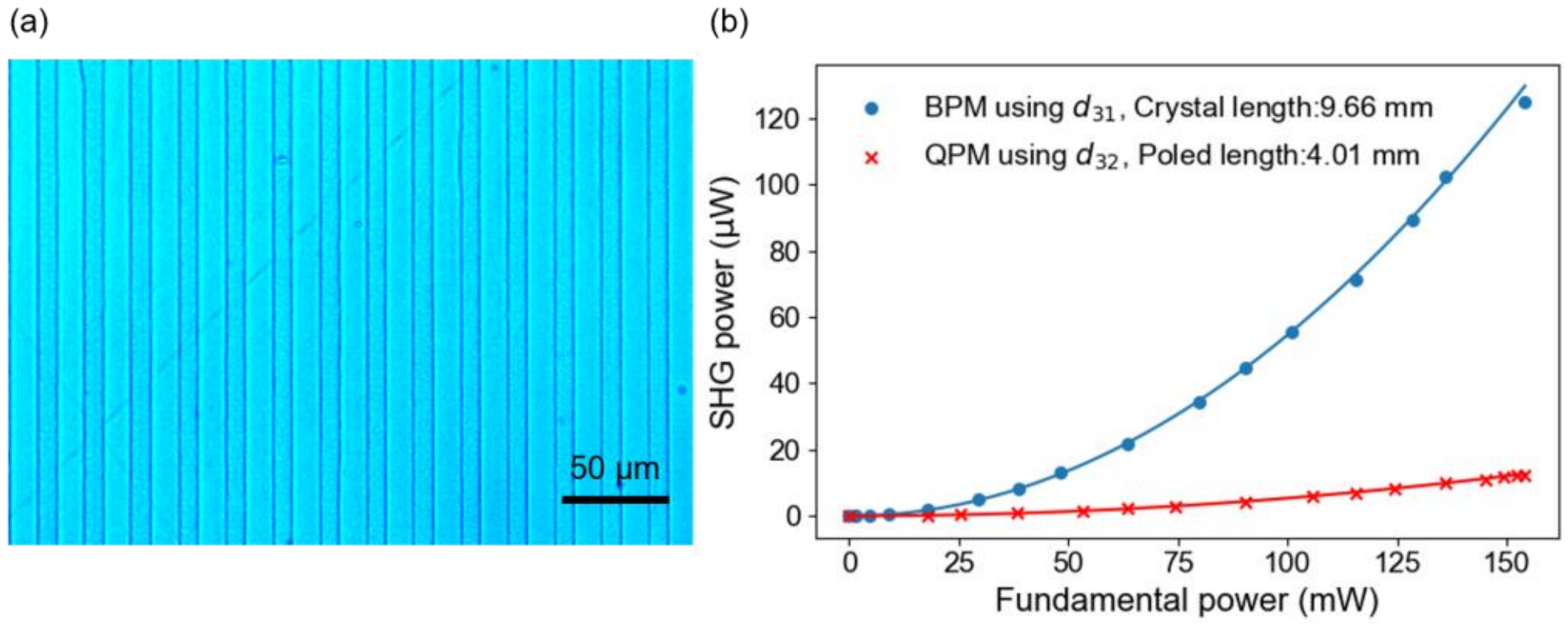

Fig. 6 (a) Microscope image of a 300-μm thick periodically-poled BMF crystal, which was selectively etched for visualizing the poled domains. The nominal poling period was 21.896 μm, corresponding to the 1st order of QPM. (b) Comparison of SHG output average power between BPM using $d_{31}$ and QPM using $d_{32}$.

## 4. Discussion

### *4.1 Comparison with previous studies*

The NLO coefficients reported in 2012 [24] are approximately one order of magnitude larger than those reported in the other studies, as well as in our work. Their Maker fringes do not agree with the theoretically expected ones. In particular, the measured fringes for $d_{31}$ shows no BPM feature. Besides, the use of $LiNbO_3$ ($d_{31}$=4.35 pm/V) as a reference to determine absolute NLO coefficients could have caused errors, because the NLO coefficients of BMF are found to be two orders of magnitude smaller than $d_{31}$ of $LiNbO_3$, making a direct comparison and scaling challenging. They state that their values are closer to the first-principles calculations, which yielded $d_{32}$=0.055 pm/V and $d_{33}$=0.190 pm/V [34]. However, only the reported $d_{33}$=0.12 pm/V is of the same order of magnitude as the calculation.

The difference in $d_{31}$ between our work and the earlier two reports in the 1970s [19,23] may be attributed to the uncertainty in the coherence length used in the analysis. Bergman *et al.* did not manage to measure it quantitatively and put it as infinity [19]. Bechthold *et al.* also overestimated the coherence length by 15% [23], which could result in 15% underestimation of $d_{31}$. In addition, the long coherence lengths make it challenging to reliably determine $d_{31}$, particularly for thin samples. The number of observable fringes is limited for thin samples, and the Maker fringes are affected by peaks arising from BPM. These factors complicate the analysis of the fringes. To mitigate these challenges, we used thick samples (>10 mm) and restricted our analysis to a narrow angular range to avoid the BPM peaks.

### *4.2 Future application of BMF for VUV nonlinear optics*

The larger $d_{31}$ value obtained in this work, although still not the largest of the five, is favorable for realizing wavelength conversion into the VUV region. Figure 7(a) shows the estimated wavelength dependence of the NLO coefficients. Assuming a constant Miller's delta [35], $d_{31}$ is roughly estimated to be 1.4 times larger, for example at a SHG wavelength of 180 nm, than the value determined at 532 nm. The use of $d_{31}$ offers one of the longest coherence lengths, or poling periods for QPM, as shown in Fig. 7(b). The smaller poling periods required at shorter wavelengths are one of the current technological obstacles in efficient VUV generation. Although the differences in coherence lengths among the corresponding coefficients decrease at shorter wavelengths (cf. Fig. 7(b)), the coherence length associated with $d_{31}$ is still 1.3 times longer than that associated with $d_{32}$ for SHG to 180 nm. The longer coherence length allows the use of a lower QPM order, relaxing the fabrication requirements while retaining the advantage of the relatively large $d_{31}$. In fact, the first demonstration of 148-nm generation using periodically-poled BMF used $d_{31}$ owing to the long poling period required [21].

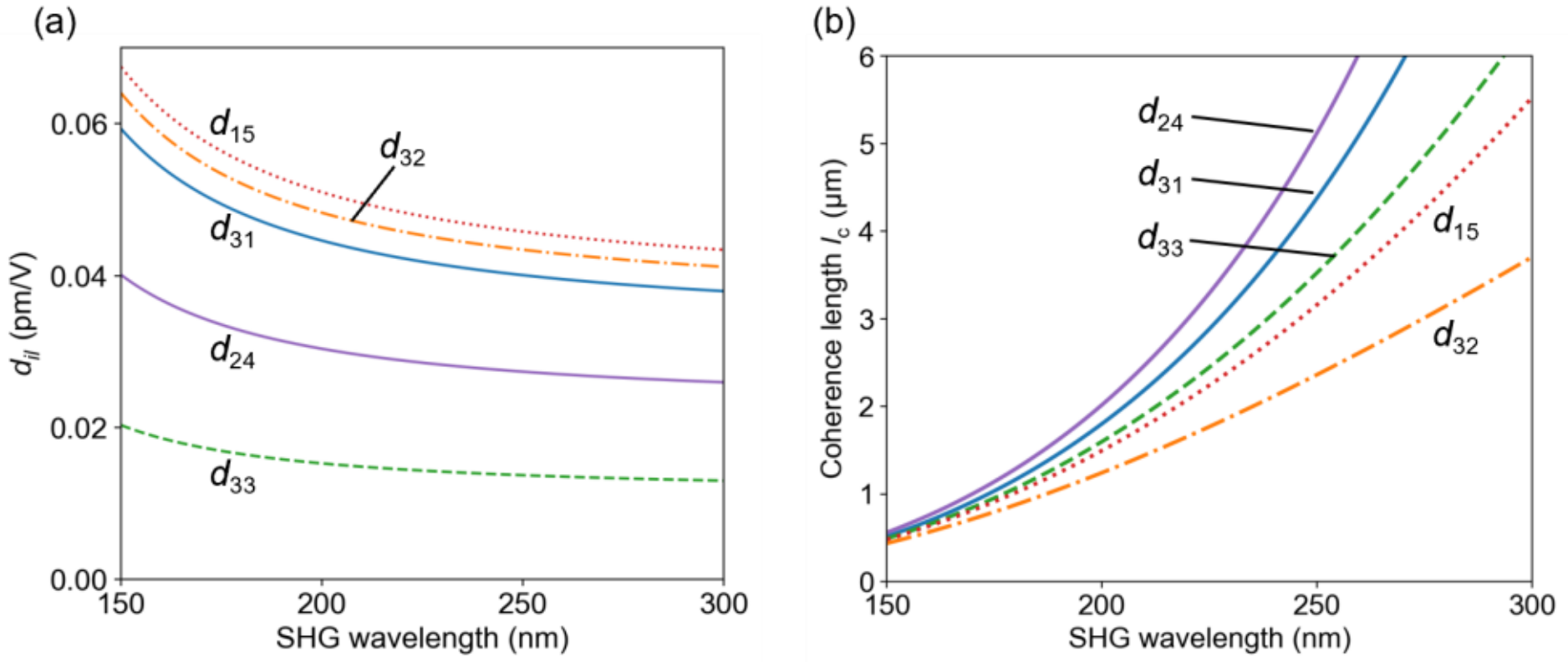

Fig. 7 (a) Changes of the NLO coefficients of BMF in the short wavelength region, assuming the constant Miller's delta. (b) Changes of the coherence lengths.

## 5. Conclusion

We redetermined the second-order NLO coefficients of BMF. Our rotational Maker fringe and wedge measurements confirmed that the NLO coefficients of BMF are in the range of 0.01–0.04 pm/V, consistent with the two previous reports from the 1970s [19,23]. Contrary to these reports, however, we suggest that $d_{31}$ and $d_{15}$ are comparable to $d_{32}$, which was previously reported to be the largest. The scale of $d_{31}$ was also supported by our single-pass SHG experiments. The key to a reliable determination of NLO coefficients would be using a reference crystal possessing an NLO coefficient comparable with investigated coefficients, ensuring a sufficient number of fringes for robust fitting, and avoiding phase-matching peaks. The long coherence length associated with $d_{31}$ allows for larger QPM periods than those required for $d_{32}$, facilitating the fabrication of lower-order QPM devices while retaining the advantage of the moderately large optical nonlinearity. To achieve high conversion efficiency in VUV generation, an additional enhancement mechanism would still be necessary, such as an external enhancement cavity [21], a whispering-gallery-mode resonator [36], or a QPM waveguide [37]. Our results provide useful guidelines for the design and fabrication of periodically-poled BMF devices for VUV generation.

## Appendix 1: Maker fringe envelope functions for orthorhombic crystals with mm2 point group

For crystals belonging to point group mm2, such as BMF, the theoretical equations for the Maker fringe measurements are given in [23]. The envelope function $f(\theta)$ in Eq. (2) is denoted as

$$f(\theta) = \frac{1024\pi^3}{c}\left(\frac{4\cos^2\theta}{w^{(\omega)}+\cos\theta}\right)^2 \frac{w^{(\omega)} n_1^{(2\omega)2}\cos\theta + v^{(2\omega)2}}{(v^{(2\omega)}-w^{(\omega)})^2(v^{(2\omega)}+w^{(\omega)})\left(v^{(2\omega)}+n_1^{(2\omega)2}\cos\theta\right)^3}$$

for $d_{31}$, $d_{32}$,

$$f(\theta) = \frac{1024\pi^3}{c}\left(\frac{4\cos^2\theta}{w^{(\omega)}+\cos\theta}\right)^2 \frac{w^{(2\omega)}(w^{(\omega)}+\cos\theta)}{(w^{(2\omega)}-w^{(\omega)})^2(w^{(2\omega)}+w^{(\omega)})(w^{(2\omega)}+\cos\theta)^3}$$

for $d_{33}$, and

$$f(\theta) = \frac{1024\pi^3}{c}\left(\frac{4v^{(\omega)}\cos^2\theta}{\left(v^{(\omega)}+n_1^{(\omega)2}\cos\theta\right)(w^{(\omega)}+\cos\theta)}\right)^2$$
$$\times \frac{w^{(2\omega)}[(v^{(\omega)}+w^{(\omega)})/2+\cos\theta]}{(w^{(2\omega)}-(v^{(\omega)}+w^{(\omega)})/2)^2(w^{(2\omega)}+(v^{(\omega)}+w^{(\omega)})/2)(w^{(2\omega)}+\cos\theta)^3}$$

for $d_{15}$, $d_{24}$. Here, $n_1$ is the refractive index for the light polarized parallel to the plane of incidence and the crystal surface. Both $v$ and $w$ are equivalent to $n\cos\theta'$. Note that $v$ is used for the s-polarized beam and $w$ for the p-polarized beam.

## Appendix 2: Dependence of the coherence lengths on wavelengths

The coherence lengths of BMF for different wavelengths calculated by Eq. (5) are shown in Fig. 8. The longest coherence length changes from $l_{c31}$ to $l_{c24}$ at SHG wavelengths shorter than 516 nm, as type-II non-critical BPM occurs at 476 nm.

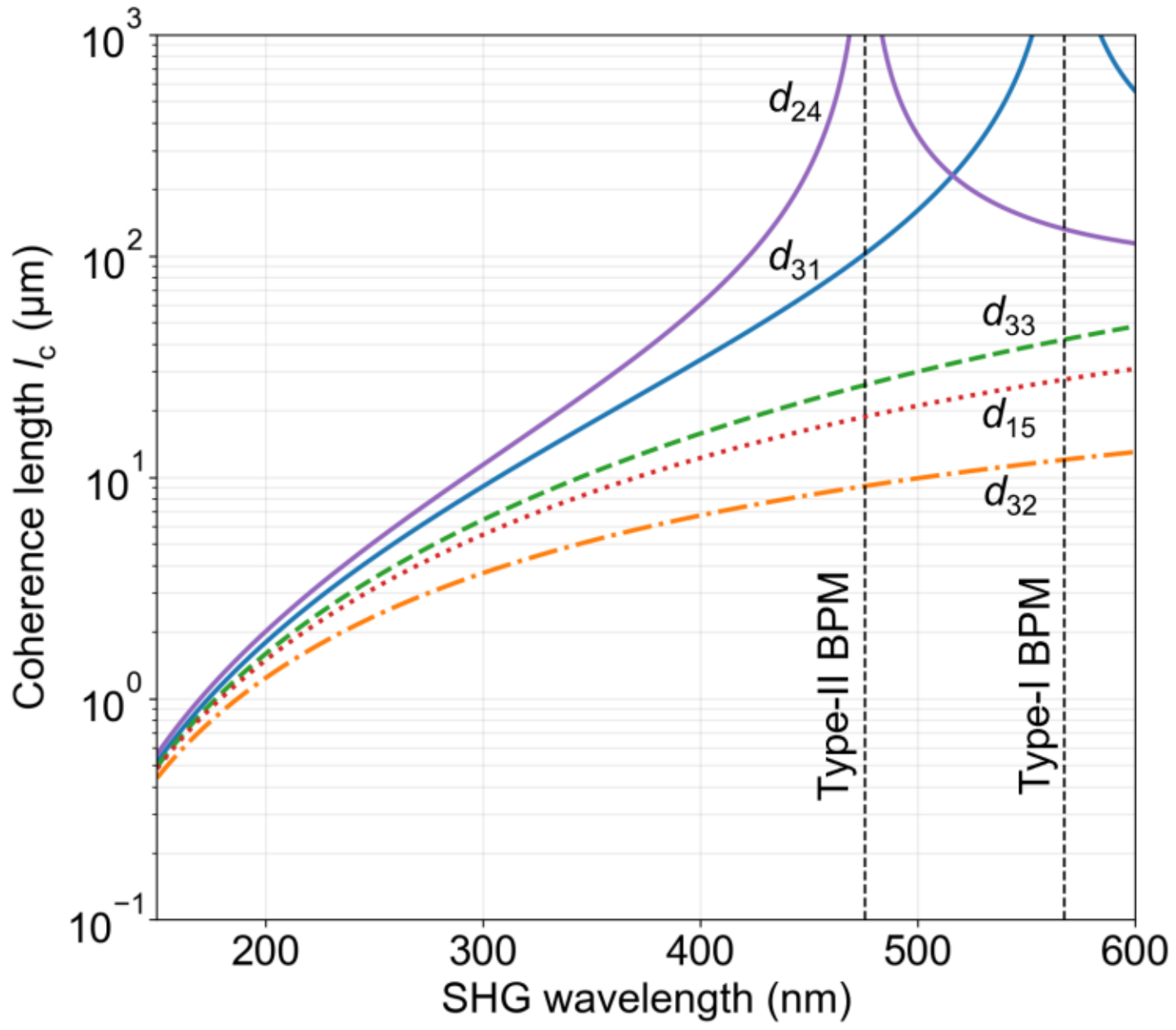

Fig. 8 Dependence of the coherence lengths on SHG wavelengths. Type-II non-critical BPM occurs at 476 nm associated with $d_{24}$, while type-I non-critical BPM occurs at 567 nm associated with $d_{31}$.

**Funding.** German Federal Ministry of Research, Technology and Space (13N16924); Leibniz Association (J183/2023).

**Acknowledgement.** We are grateful to Prof. Ichiro Shoji (Chuo University, Japan) for valuable discussions.

**Disclosures.** The authors declare no conflicts of interest.

**Data availability.** Data underlying the results presented in this paper are not publicly available at this time but may be obtained from the authors upon reasonable request.

## References

1. Z. Zhang, Y. Ma, C. Niu, K. Li, K. Wang, C. Wu, and Y. Yu, "Development of all-solid-state ultraviolet lasers," J. Laser Appl. **36**, 041202 (2024).
2. I. K. Lednev, V. V. Ermolenkov, W. He, and M. Xu, "Deep-UV Raman spectrometer tunable between 193 and 205 nm for structural characterization of proteins," Anal. Bioanal. Chem. **381**, 431–437 (2005).
3. F. Fan, Z. Feng, and C. Li, "UV Raman spectroscopic study on the synthesis mechanism and assembly of molecular sieves," Chem. Soc. Rev. (2010).
4. Q.-J. Peng, N. Zong, S.-J. Zhang, Z.-M. Wang, F. Yang, F.-F. Zhang, Z.-Y. Xu, and X.-J. Zhou, "DUV/VUV all-solid-state lasers: Twenty years of progress and the future," IEEE J. Sel. Top. Quantum Electron. **24**, 1–12 (2018).
5. D. R. Leibrandt, S. G. Porsev, C. Cheung, and M. S. Safronova, "Prospects of a thousand-ion $Sn^{2+}$ Coulomb-crystal clock with sub-$10^{-19}$ inaccuracy," Nat. Commun. **15**, 5663 (2024).
6. E. Peik and C. Tamm, "Nuclear laser spectroscopy of the 3.5 eV transition in Th-229," Europhys. Lett. **61**, 181–186 (2003).
7. V. Petrov, F. Noack, D. Shen, F. Pan, G. Shen, X. Wang, R. Komatsu, and V. Alex, "Application of the nonlinear crystal $SrB_4O_7$ for ultrafast diagnostics converting to wavelengths as short as 125 nm," Opt. Lett. **29**, 373–375 (2004).
8. A. I. Zaitsev, A. S. Aleksandrovsky, A. D. Vasiliev, and A. V. Zamkov, "Domain structure in strontium tetraborate single crystal," J. Cryst. Growth **310**, 1–4 (2008).
9. P. Trabs, F. Noack, A. S. Aleksandrovsky, A. I. Zaitsev, and V. Petrov, "Generation of coherent radiation in the vacuum ultraviolet using randomly quasi-phase-matched strontium tetraborate," Optics letters **41**, 618–621 (2016).
10. V. Lal, M. V. Okhapkin, J. Tiedau, N. Irwin, V. Petrov, and E. Peik, "Continuous-wave laser source at the 148 nm nuclear transition of Th-229," Optica **12**, 1971 (2025).

11. C. Chen, G. Wang, X. Wang, Y. Zhu, Z. Xu, T. Kanai, and S. Watanabe, "Improved sellmeier equations and phase-matching characteristics in deep-ultraviolet region of $KBe_2BO_3F_2$ crystal," IEEE J. Quantum Electron. **44**, 617–621 (2008).
12. S.-B. Dai, M. Chen, S.-J. Zhang, Z.-M. Wang, F.-F. Zhang, F. Yang, Z.-C. Wang, N. Zong, L.-J. Liu, X.-Y. Wang, J.-Y. Zhang, Y. Bo, D.-F. Cui, Q.-J. Peng, R.-K. Li, C.-T. Chen, and Z.-Y. Xu, "2.14 mW deep-ultraviolet laser at 165 nm by eighth-harmonic generation of a 1319 nm Nd:YAG laser in KBBF," Laser Phys. Lett. **13**, 035401 (2016).
13. T. Nakazato, I. Ito, Y. Kobayashi, X. Wang, C. Chen, and S. Watanabe, "Phase-matched frequency conversion below 150 nm in $KBe_2BO_3F_2$," Opt. Express **24**, 17149–17158 (2016).
14. M. Scholz, D. Opalevs, P. Leisching, W. Kaenders, G. Wang, X. Wang, R. Li, and C. Chen, "A bright continuous-wave laser source at 193 nm," Appl. Phys. Lett. **103**, 051114 (2013).
15. H. Su, Z. Yan, X. Hou, and M. Zhang, "Fluorooxoborates: A precious treasure of deep-ultraviolet nonlinear optical materials," Jiegou Huaxue **42**, 100027 (2023).
16. F. Zhang, Z. Chen, C. Cui, Z. Yang, M. Mutailipu, F. Li, X. Hou, X. Long, and S. Pan, "Vacuum ultraviolet second-harmonic generation in $NH_4B_4O_6F$ crystal," Nature **650**, 97–101 (2026).
17. K. Shimamura, E. G. Víllora, H. Zeng, M. Nakamura, S. Takekawa, and K. Kitamura, "Ferroelectric properties and poling of $BaMgF_4$ for ultraviolet all solid-state lasers," Appl. Phys. Lett. **89**, 8–11 (2006).
18. E. Víllora, K. Shimamura, K. Sumiya, and H. Ishibashi, "Birefringent- and quasi phase-matching with $BaMgF_4$ for vacuum-UV/UV and mid-IR all solid-state lasers," Opt. Express **17**, 12362–12378 (2009).
19. J. G. Bergman, G. R. Crane, and H. Guggenheim, "Linear and nonlinear optical properties of ferroelectric $BaMgF_4$ and $BaZnF_4$," J. Appl. Phys. **46**, 4645–4646 (1975).
20. S. C. Buchter, T. Y. Fan, V. Liberman, J. J. Zayhowski, M. Rothschild, E. J. Mason, A. Cassanho, H. P. Jenssen, and J. H. Burnett, "Periodically poled $BaMgF_4$ for ultraviolet frequency generation," Opt. Lett. **26**, 1693 (2001).
21. K. Subramanian, H. Tanaka, S. J. Herr, N. K. Sah, G. Jha, F. Zacherl, S. A. Pradeep, V. Andriushkov, K. Zhang, D. Fenner, Y. Wang, M. Hugenschmidt, F. Kühnemann, G. G. M. Bonetti, S. Ui, M. Bickermann, C. Ma, X. Zheng, M. Zopf, B. Lommel, J. C. Müller, S. Hannig, D. Budker, F. Schmidt-Kaler, and L. von der Wense, "Generation of continuous-wave laser light at 148.4 nm using cavity-enhanced second-harmonic generation in $BaMgF_4$," Optica **13**, 2089 (2026).
22. F. Gingl, "$BaMgF_4$ and $Ba_2Mg_3F_{10}$: New examples for structural relationships between hydrides and fluorides," Z. Anorg. Allg. Chem. **623**, 705–709 (1997).
23. P. S. Bechthold and S. Haussühl, "Nonlinear optical properties of orthorhombic barium formate and magnesium barium fluoride," Appl. Phys. **14**, 403–410 (1977).
24. J. Chen, X. Chen, Y. Ma, Y. Zheng, A. Wu, H. Li, L. Jiang, and J. Xu, "Measurement of second-order nonlinear optical coefficients of $BaMgF_4$," J. Opt. Soc. Am. B **29**, 665 (2012).
25. I. Shoji, T. Kondo, A. Kitamoto, M. Shirane, and R. Ito, "Absolute scale of second-order nonlinear-optical coefficients," JOSA B **14**, 2268–2294 (1997).
26. S. Herr, H. Tanaka, I. Breunig, M. Bickermann, and F. Kuehnemann, "Fanout periodic poling of $BaMgF_4$ crystals," Opt. Mater. Express **13**, 2158–2164 (2023).
27. J. Jerphagnon and S. K. Kurtz, "Maker fringes: A detailed comparison of theory and experiment for isotropic and uniaxial crystals," J. Appl. Phys. **41**, 1667–1681 (1970).
28. J. Jerphagnon and S. K. Kurtz, "Optical nonlinear susceptibilities: Accurate relative values for quartz, ammonium dihydrogen phosphate, and potassium dihydrogen phosphate," Phys. Rev. **1**, 1739–1744 (1970).
29. M. Braun, F. Bauer, T. Vogtmann, and M. Schwoerer, "Precise second-harmonic generation Maker fringe measurements in single crystals of the diacetylene NP/4-MPU and evaluation by a second-harmonic generation theory in $4 \times 4$ matrix formulation and ray tracing," J. Opt. Soc. Am. B, JOSAB **14**, 1699–1706 (1997).
30. M. Abe, I. Shoji, J. Suda, and T. Kondo, "Comprehensive analysis of multiple-reflection effects on rotational Maker-fringe experiments," J. Opt. Soc. Am. B, JOSAB **25**, 1616–1624 (2008).
31. I. Shoji, T. Kondo, and R. Ito, "Second-order nonlinear susceptibilities of various dielectric and semiconductor materials," Opt. Quantum Electron. **34**, 797–833 (2002).
32. G. D. Boyd and D. A. Kleinman, "Parametric interaction of focused Gaussian light beams," J. Appl. Phys. **39**, 3597–3639 (1968).
33. M. M. Fejer, G. A. Magel, D. H. Jundt, and R. L. Byer, "Quasi-phase-matched second harmonic generation: tuning and tolerances," IEEE J. Quantum Electron. **28**, 2631–2654 (1992).
34. H. Huang, Z. S. Lin, L. Bai, R. He, and C. T. Chen, "Mechanism of the linear and nonlinear optical effects of $SrAlF_5$ and $BaMgF_4$ crystals," Solid State Commun. (2010).
35. R. C. Miller, "Optical second harmonic generation in piezoelectric crystals," Appl. Phys. Lett. **5**, 17–19 (1964).
36. A. Mrokon, H. Kraft, D. Shin, H. Tanaka, S. J. Herr, K. Buse, and I. Breunig, "Electrically induced resonance shifts of whispering gallery resonators made of barium magnesium fluoride," APL Photonics **10**, (2025).
37. C. Wang, C. Langrock, A. Marandi, M. Jankowski, M. Zhang, B. Desiatov, and Lončar, "Ultrahigh-efficiency wavelength conversion in nanophotonic periodically poled lithium niobate waveguides," Optica **5**, 1438–1441 (2018).